# Influence of $P_2O_5$ and $SiO_2$ addition on the phase, microstructure, and electrical properties of $KNbO_3$

**S. Ullah[1,2,*] · I. Ullah[1] · Y. Iqbal[1] · A. Manan[1,3] · S. Ali[1] · A. Khan[4]**

**Abstract** In this contribution, the effect of $P_2O_5$ and $SiO_2$ addition on the phase, microstructure, and electrical properties of $KNbO_3$ was studied. Sample powders with the general formula (1-x)$KNbO_3$.x$P_2O_5$ (x = 0.03, 0.05) and (1-x)$KNbO_3$.x$SiO_2$ (x = 0.1) were prepared via mixed-oxide route. The thermal behavior of the mixed-milled powder was investigated by TG/DTA which revealed an overall weight loss of 33.4 wt % in the temperature range of $30 \leq T \leq 1200$ °C and crystallization exotherm occurring at about 795 °C. The present results indicated that $P_2O_5$ acted as a sintering aid and lowered the sintering temperature by about 30 °C and promoted densification of $KNbO_3$. Sample compositions at various stages of processing were characterized using X-ray diffraction. Samples sintered at $T \leq 1020$ °C revealed mainly $KNbO_3$ together with a couple of low-intensity $K_3NbO_4$ peaks as a secondary phase. The SEM images of (1-x)$KNbO_3$.x$SiO_2$ (x = 0.1) samples showed a slight increase in the average grain size from 3.76 $\mu$m to 3.86 $\mu$m with an increase in sintering temperature from 1000 °C to 1020 °C. Strong variations in dielectric constant and loss tangent were observed due to $P_2O_5$ and $SiO_2$ addition as well as frequency of the applied AC signals.

Correspondence author: Dr. Saeed Ullah
E-mail: saeedullah.phy@gmail.com
Tel.: +92-342-9070068

1 Materials Research Laboratory, Department of Physics, University of Peshawar, 25120, Pakistan

2 Department of Physics, Gomal University, Dera Ismail Khan 29220, KP, Pakistan

3 Department of Physics, University of Science and Technology, Bannu Khyber Pakhtunkhwa, Pakistan

4 Center of Excellence in Solid State Physics, University of the Punjab, Lahore-54590, Pakistan

## 1 Introduction

To meet the massive demands of the evolving society advanced technological developments are required[1]. Therefore, it is highly desirable to use new and advanced materials with improved properties[2,3]. Currently, the lead-based piezoelectric materials, such as Pb(Zr,Ti)$O_3$ (PZT) and Pb($Mg_{0.33}Nb_{0.67}$)$O_3$-$PbTiO_3$ (PMN-PT), dominate the commercial piezoelectric applications. While such materials, due their superior properties, have been widely used in microelectronics and electronic devices as dielectric capacitors, ultrasonic or medical transducers, ultrasonic motors, and sensors[4,5]. However, one of the major drawbacks of these materials is the environmental and health-related issues. These materials contain about 70 wt% of poisonous lead-oxide which is an ecological hazard[6]. Material scientists are looking forward to orient the current research towards environmentally friendly materials of comparable properties[7,8]. Recently, a number of materials such as barium titanate, bismuth-alkaline metal titanates, and alkaline-niobates are revisited as potential alternatives to lead-based piezoceramics[9].

Among those alternatives, potassium niobate [$KNbO_3$] (KN) based materials are the front-running candidates due to their large non linear optical[10] and electro-optic coefficient[11], moderate dielectric constant, and excellent photorefractive and electrical properties[12,13,14,15,16]. Additionally, such materials also offer possibilities for applications, for example, in the production of surface acoustic wave (SAW) devices and wide-band SAW filters[17,18]. However, the den-sification of these materials is problematic due to several reasons. One of the main reason is the phase stability of $KNbO_3$ which is limited to 1040 °C which further restrains the high-temperature sintering[19]. Additionally, potassium oxide is volatile and evaporates easily with rising temperature during the sintering process and hence leads to the formation of some unstable secondary phases in the final product. Furthermore, slight changes in the stoichiometry can also result in the formation of unwanted phases.

Various techniques have been developed for the improvement of densification of niobate-based ceramics, for example, hot pressing and spark plasma sintering (SPS). However, these techniques are not suitable to enhance the sintering in various commercial applications[20,21]. Based on previous efforts, it has been concluded that the modification of stoichiometry[15] or addition of sintering agents such as CuO, $CeO_2$, $MnO_2$, $TiO_2$ and other oxides, have a significant impact on the improvement of sinterability[22,23,24,25,26,27,28,29]. These oxides form a liquid phase and promote densification[25,26,28].

The objective of the present work was to investigate the effect of $P_2O_5$ and $SiO_2$ addition on the phase, microstructure and electrical properties of KN. A number of experimental groups have reported the addition of various oxides into KN-based materials. However, to our knowledge, no report regarding the use of $P_2O_5$ and $SiO_2$ addition has, to date been appeared in the literature.

## 2 Materials and Experiment

The (1-x)$KNbO_3$.x$P_2O_5$ (x = 0.03, 0.05) and (1-x)$KNbO_3$.x$SiO_2$ (x = 0.1) samples were prepared via a conventional mixed-oxide route using $K_2CO_3$ and $Nb_2O_5$ as starting materials. The amount of oxides, to produce the stoichiometric KN, were calculated on the basis of their molecular weights. The weighed amounts were wet ball milled for 24 h in disposable polyethylene mill jars using Y-toughened $ZrO_2$ balls as grinding media and iso-propanol as a lubricant to make the slurry. The slurry was then dried overnight in an oven at 95 °C. To determine the phase transformation temperatures, thermal gravimetry and differential thermal analysis (TG/DTA) was carried out in the temperature range of 30-1200 °C with a heating/cooling rate of 5 °C/min using a Perkin Elmer TG/DTA unit. The mix-milled powder was then calcined at 850 °C for 2 h at the same heating/cooling rate. To lower the sintering temperature, the as-calcined powder were then mixed with 0.03 and 0.05 wt% $P_2O_5$ and 0.1 wt% $SiO_2$ and re-milled for 24 h.

The powders were then pressed into pellets of 10 mm diameters at ∼100 MPa using a manual pellet press, followed by sintering at various temperatures. The apparent densities of the specimens were measured employing Archimedes method, however, due to its hygroscopic nature, the $P_2O_5$-added samples dissolved in water[30]. Densities were then calculated from the mass and dimensions of the pellets and converted to the relative densities using theoretical density (TD) of $KNbO_3$ (4.167 g/cm$^3$)[26].

The crystalline phases of compositions, at different stages of processing, were assessed by a JEOL X-ray diffractometer (model JDX-3532) in the $2\theta$ range of 10-70°with Cu $K_\alpha$ radiations operating at 40 kV and 30 mA with a step size of 0.02°. The sintered samples were cut with a diamond fine cutting saw, smoothened using a polishing wheel with sandpaper and then polished using a polishing wheel with a nylon cloth and diamond paste. After polishing, the samples were thermally etched at a temperature 10% less than the corresponding sintering temperatures. Polished samples were then gold-coated to avoid charging in the scanning electron microscope (SEM). Microstructural characterization was performed using SEM (JEOL JSM 5910). The grain size measurements were performed using mean intercept length method at different areas of the sample. The dielectric constant ($\varepsilon_r$) and loss tangent (tan$\delta$) were measured in a frequency range of 1 MHz to 1 GHz employing a 4287A LCR meter. All the experiments were carried out using the equipment installed at the Materials Research Laboratory (MRL) and Centralized Resource Laboratory (CRL), University of Peshawar.

## 3 Results and Discussions

### 3.1 TG/DTA Analysis

Fig. 1 depicts the TG/DTA curve of the stoichiometric precursor of $KNbO_3$ showing three major and two minor downward slopes in the employed temperature range i.e. from room temperature to 1200 °C. Consistent with the TG slopes, four endotherms were observed on the DTA curve at temperatures 52-153 °C, 464 °C, 792 °C and 1126 °C respectively. The wt% losses determined at 50-171 °C, 430-475 °C and 475-900 °C were 13.23 wt%, 4.28 wt%, and 11.11 wt% respectively. Furthermore, relatively minor wt% losses (i.e., 3.75 wt% and 1.052 wt%) were observed at 171-430 °C and 900-1200 °C. The observed weight losses at different temperatures are listed in Table 1. The overall wt% loss was 33.4 wt%. The wt% loss at 50-171 °C may be due to the dehydration of the sample, i.e., the removal of surficial water present in the raw materials[31].

The weight loss observed in the temperature range 430-475 °C and the observed DTA endotherm at 464 °C may be associated with the glass transition of some of the glass formers (i.e., $P_2O_5$ and $SiO_2$) present in the batch. In the temperature range of 475-900 °C, the observed weight loss is attributed to the decomposition of $K_2CO_3$, releasing $CO_2$

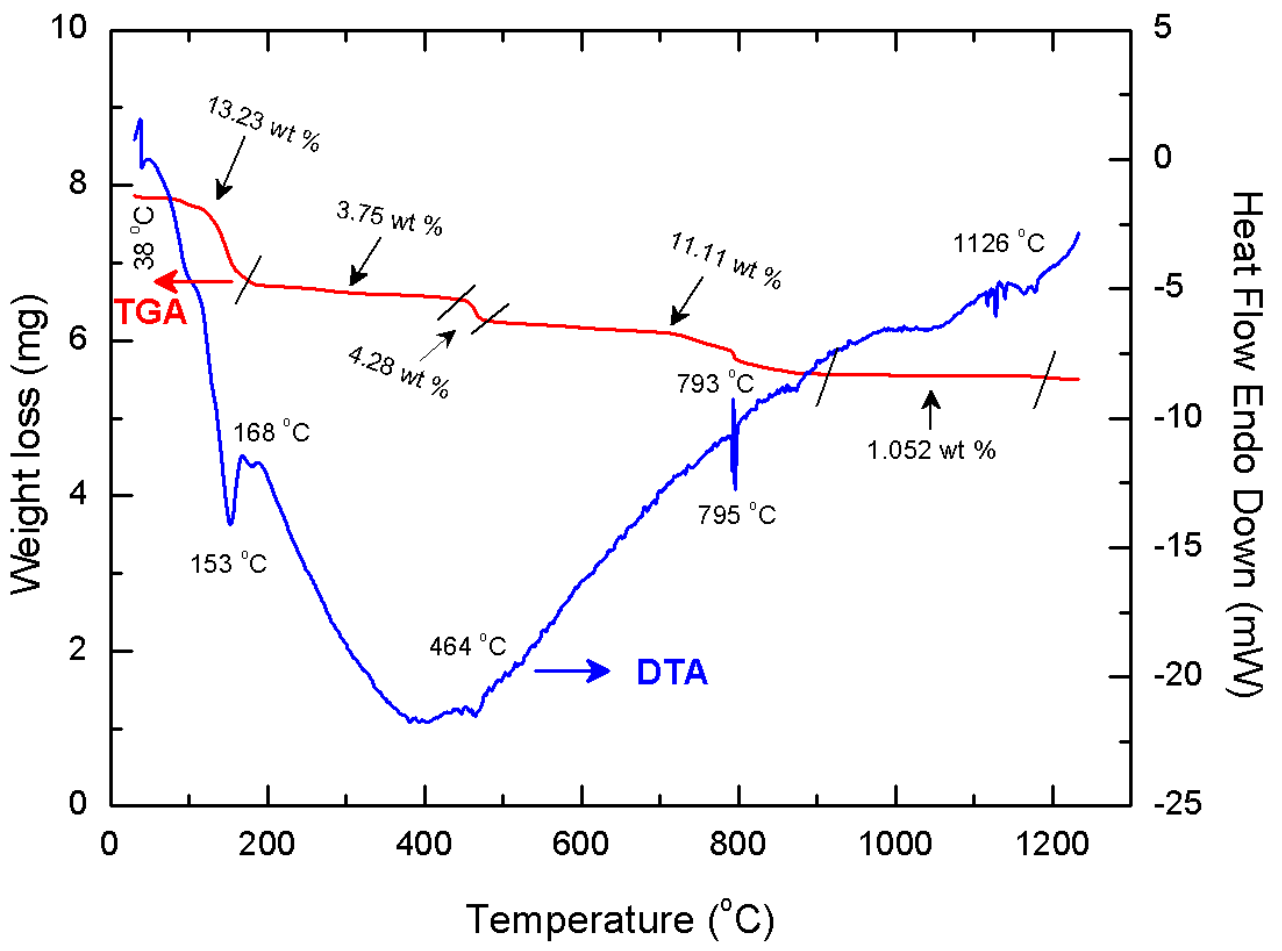

**Fig. 1** TG/DTA curves of as-mixed-milled material, showing major weight losses in the temperature ranges of 50-171 °C, 430-475 °C and 475-900 °C with the corresponding endotherms at 153 °C, 464 °C and 792 °C. Note the corresponding endo-or exo-therms on the DTA curve.

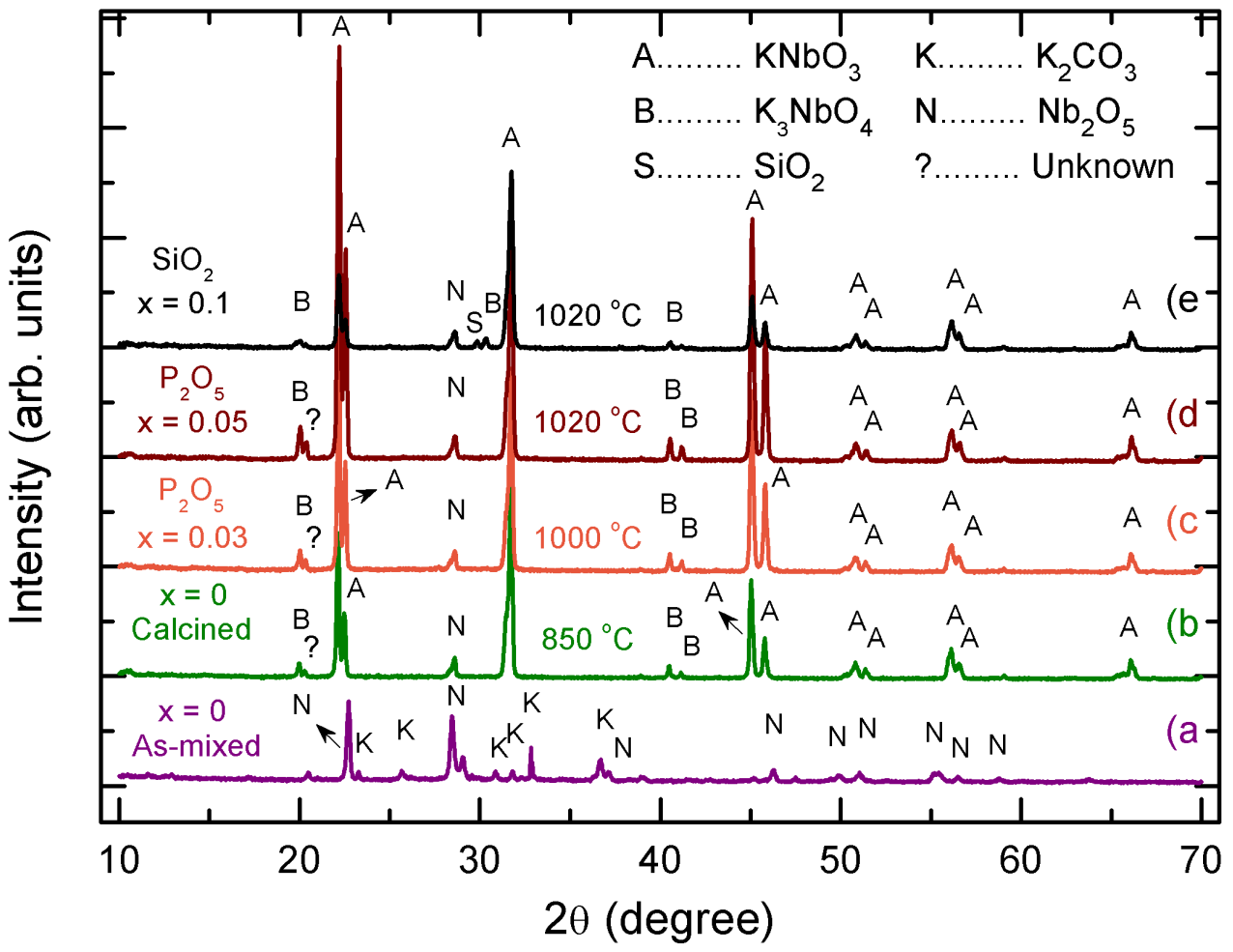

**Fig. 2** X-ray diffraction patterns recoded from a) as-mix-milled batch b) (1-x)$KNbO_3$.xP2O5 (x = 0) calcined at 850 °C c) the (1-x)$KNbO_3$.x$P_2O_5$ (x = 0.03) sintered at 1000 °C d) the (1-x)$KNbO_3$.x$P_2O_5$ (x = 0.05) sintered at 1020 °C and e) the (1-x)$KNbO_3$.x$SiO_2$ (x = 0.1) sintered at 1020 °C.

by combustion reaction (See Eq. 1).

$$K_2CO_3 \longrightarrow K_2O + CO_2 \quad (1)$$

A corresponding endotherm was observed on the DTA curve at 792 °C. In the temperature range 900-1200 °C, the relatively small wt% loss is related to the removal of entrapped gases upon partial melting of the sample releasing $CO_2$ and other gases[32]. In the temperature range 900-1200 °C, the DTA curve showed a corresponding endotherm at 1126 °C consistent with the TG results. The exotherm observed on the DTA curve at 795 °C is associated with the crystallization of potassium niobate[33] as confirmed by phase analysis (See Fig. 2). Based on the TG/DTA analysis, temperatures in the range from 800 to 1030 °C were selected for calcinations. The mixed-milled sample was calcined in air at 850 °C for 2 h with a heating/cooling rate 5 °C /min.

**Table 1** Weight loss of $KNbO_3$ at various temperatures.

| S. No | Temperature range | Weight loss (mg) | Weight loss (%) |
|---|---|---|---|
| 1 | 50 - 171 °C | 1.04 | 13.23 |
| 2 | 171 - 430 °C | 0.255 | 3.75 |
| 3 | 430 - 475 °C | 0.280 | 4.28 |
| 4 | 475 - 900 °C | 0.696 | 11.11 |
| 5 | 900 - 1200 °C | 0.059 | 1.05 |

## 3.2 Phase Analysis

Fig. 2 shows the X-ray diffraction patterns of the as-mix-milled, calcined and sintered samples recorded at room temperature in the $2\theta$ range of 10-70°. The d-spacings and relative intensities corresponding to the XRD peaks from the as-mix-milled sample matched PDF # 70-292 for $K_2CO_3$ and PDF # 27-1312 for $Nb_2O_5$ labeled as "K" and "N" respectively which confirmed the presence of initial ingredients used in the present study. The XRD pattern of the (1-x)$KNbO_3$.x$P_2O_5$ (x = 0) sample, calcined at 850 °C for 2 h, is shown in Fig. 2(b). The d-spacings and relative intensities corresponding to major XRD peaks matched PDF # 32-0822 for $KNbO_3$ labeled as "A" which revealed the formation of $KNbO_3$ as a major phase; however, a couple of low-intensity peaks labeled as "K" and "N" matched PDF # 70-292 for $K_2CO_3$ and PDF # 27-1312 for $Nb_2O_5$ indicating incomplete reaction.

Fig. 2(c) and (d) show XRD patterns recorded from the (1-x)$KNbO_3$.x$P_2O_5$ (x = 0.03, x = 0.05) samples sintered at 1000 °C and 1020 °C respectively. The d-spacings and relative intensities corresponding to the major XRD peaks from (1-x)$KNbO_3$.x$P_2O_5$ (x = 0.03) compositions sintered at 1000 °C and (1-x)$KNbO_3$.x$P_2O_5$ (x = 0.05) sintered at 1020 °C matched the PDF # 32-0822 for $KNbO_3$ labeled as "A" which confirmed the occurrence of chemical reaction and hence the formation of $KNbO_3$ phase. A number of low-intensity peaks marked as "?" were also observed which could not be identified[26,27,28,29].

Fig. 2(e) shows the XRD pattern recorded from the (1-x)$KNbO_3$.x$SiO_2$ (x = 0.1) sample sintered at 1020 °C. The corresponding d-spacings and relative intensities of XRD peaks matched the PDF # 32-0822 for $KNbO_3$ which confirmed the presence of $KNbO_3$ phase. For this composition, the unknown peak disappeared, however, an additional low-intensity peak marked as "S" appeared at $2\theta = 29°$ which matched the PDF # 50-1431 for $SiO_2$. Additionally, the thermal treatment processes (such as calcination and sintering) caused the emergence of secondary phases labeled as ”B” (See Fig. 2(b-e)). The appearance of these peaks can be assigned to $K_3NbO_4$ (PDF # 52-1894) originating from the slight changes in the stoichiometry due to volatile alkaline oxides during the sintering process.

### 3.3 Density measurement

In the present work, the density of 3.07 g/cm$^3$ was measured for the as-mixed-milled sample, which is 73.6 % of the theoretical density of $KNbO_3$[26]. The addition of 0.03 and 0.05 wt% $P_2O_5$ resulted in the densification of the material and hence enhanced the density by about 11.35 and 14.55 % of that of the mixed-milled sample. In addition, to increasing x value, the observed increase in density was caused by the rise of sintering temperature from 1000 to 1020 °C. Furthermore, the addition of 0.1 wt% $SiO_2$ lowered the density to 2.8 g/cm$^3$ (66.04 % of TD) which is reflected from the increase in grain size (See Sec. 3.4).

### 3.4 Microstructural Analysis

Fig. 3 depicts the secondary electron SEM images (SEIs) recorded from the thermally etched, gold-coated surfaces of (1-x)$KNbO_3$.x$SiO_2$ (x = 0.1). The microstructure comprised cubic and cuboidal grains with an average grain size of 3.76 $\mu$m and 3.86 $\mu$m along with agglomerates of fine grains in samples sintered at 1000 °C and 1020 °C for 2 h, respectively. Zhou et *al.*[34] reported that pure $KNbO_3$ has an average grain size of 2.5 $\mu$m. Thus, the addition of $SiO_2$ caused an increase in the grain size of $KNbO_3$ ceramics which is evidenced by the reduction of the apparent density of the sample as commented above.

The composition analysis was carried out using energy dispersive X-ray spectroscopy (EDS). EDS detected the presence of K and Nb only in cubic and cuboidal grains with no Si. The agglomerates of fine grains contained K, Nb, and a small amount of Si. The EDS analysis (see table 2) detected 29.47 wt% K and 70.53 wt% Nb in the cubic grain labeled as ”A”, and 29.29 K wt% and 71.71 wt% Nb in cuboidal grains labeled as ”C” (see Fig. 3(a,b)). Additionally, the agglomerates of fine grains marked as ”B” revealed 29.29 wt% K, 69.09 wt% Nb and 1.62 wt% Si. In Fig 4(c,d), EDS detected 29.33 wt% K and 70.66 wt% Nb in the cubic grain labeled as ”D”, 29.87 wt% K and 71.13 wt% Nb in cuboidal grains labeled as ”F” and 28.62 wt% K, 65.28 wt% Nb and 6.10 wt% Si in agglomerates of fine grains labeled as ”E”. Furthermore, the EDS analysis of the white spot on regular shaped grain, marked by an arrow in Fig. 3(b), revealed 30.51 wt% K and 69.49 wt% Nb. The observed compositions in EDS analysis are consistent with XRD finding. The edges of the grains containing K, Nb, and Si appeared less sharp or diffused in comparison to those containing K and Nb only, indicating the inhomogeneous mixing of Si in the ceramic body.

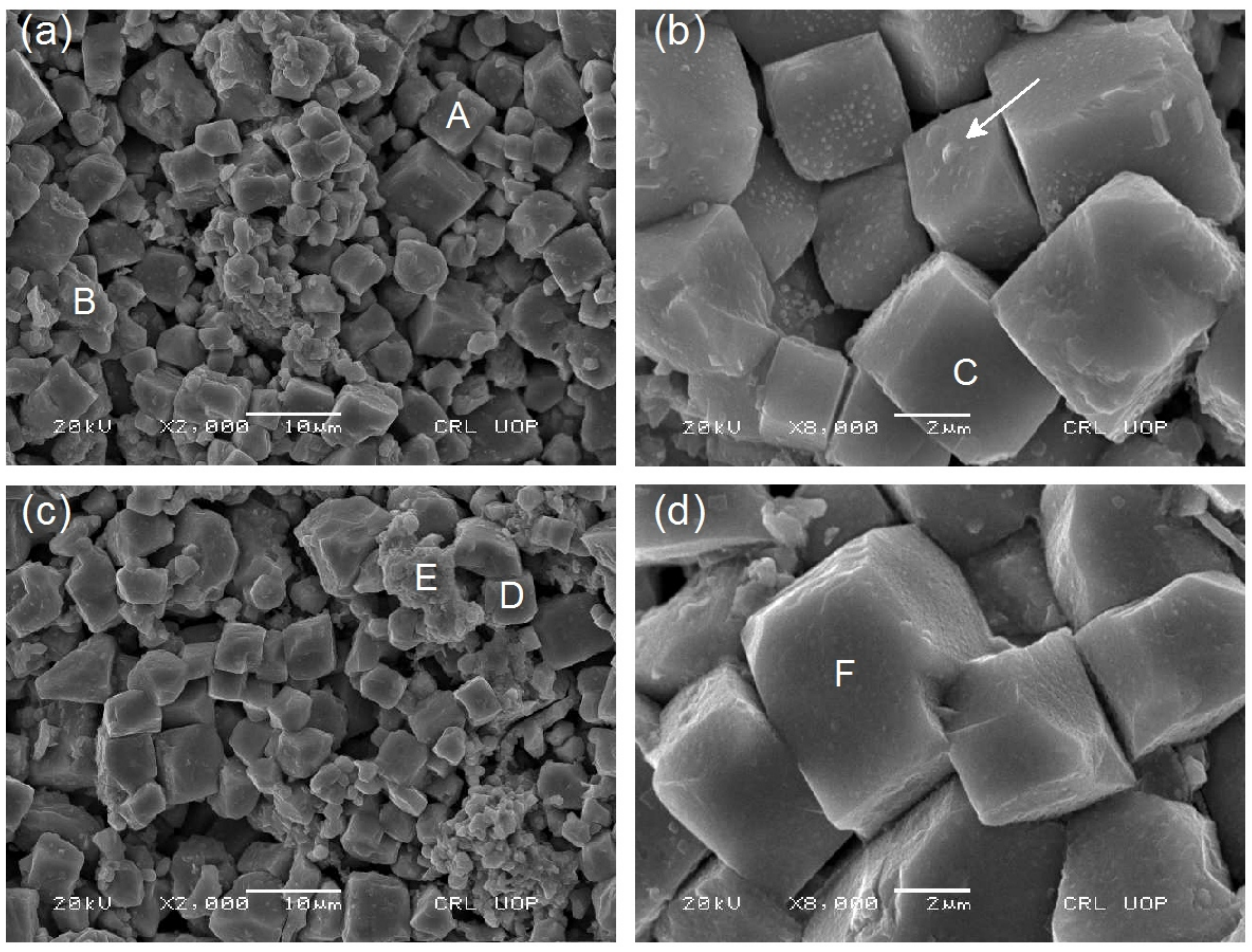

**Fig. 3** Scanning electron micrographs of thermally etched gold-coated surface of (1-x)$KNbO_3$.x$SiO_2$ (x = 0.1) (a,b) sintered at 1000°C, showing the general microstructure grains of cubic and cuboidal morphology with an average grain size of 3.76 $\mu$ m along with agglomerates of fine grains. (c,d) sintered at 1020°C, showing same morphology with an average grain size of 3.86 $\mu$ m together with agglomerates of fine grains.

**Table 2** EDS-detected compositions of various grains.

| Grains | Potassium (K %) | Niobium (Nb %) | Silicon (Si %) |
|---|---|---|---|
| A | 29.47 | 70.53 | - |
| B | 29.29 | 69.09 | 1.62 |
| C | 29.29 | 71.71 | - |
| D | 29.33 | 70.67 | - |
| E | 28.62 | 65.28 | 6.10 |
| F | 29.87 | 71.13 | - |

### 3.5 Electrical properties

Variation in $\varepsilon_r$ and tan$\delta$ as a function of frequency for the (1-x)$KNbO_3$.x$P_2O_5$ (x = 0.03, 0.05) and (1-x)$KNbO_3$.x$SiO_2$ (x = 0.1) samples, investigated at room temperature, are shown in Fig. 4. The dielectric constants and tan$\delta$ were observed to decrease with an increase in frequency as well as concentration of the sintering aids. This behavior can be attributed to various polarization effects. The observed $\varepsilon_r$ decreased drastically with an increase in frequency while the tan$\delta$ first decreased and then increased abruptly with increasing frequency, indicating that the material behaved like a conductor at high frequencies[35]. $KNbO_3$ single crystals are known to exhibit a dielectric constant of 137 and correspond to (111) and (100) planes, i.e., parallel to c-axis of its tetragonal unit cell[36]. In comparison to pure $KNbO_3$ single crystals, the

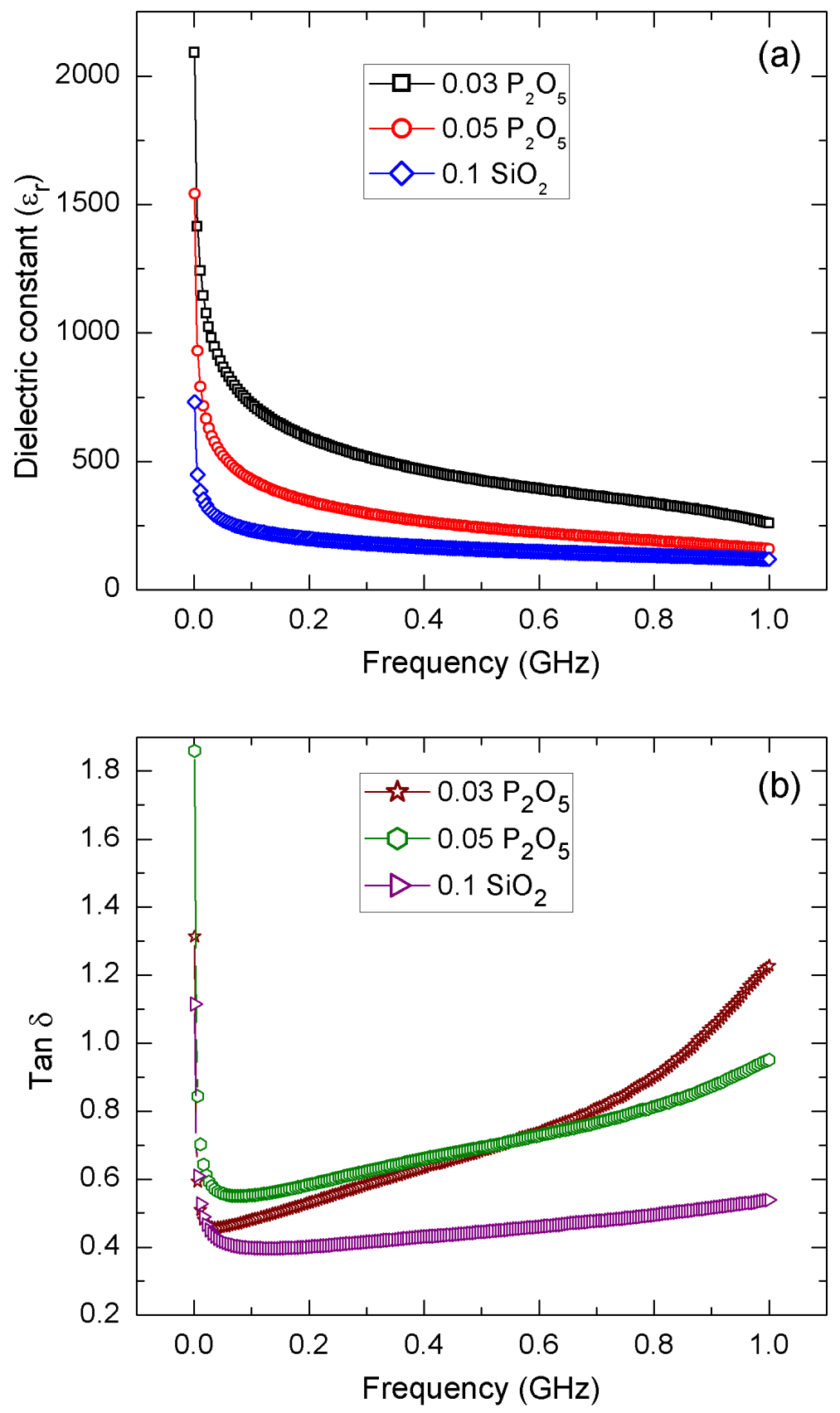

**Fig. 4** (a) $\varepsilon_r$ and (b) tan$\delta$ measured for (1-x)$KNbO_3$.x$P_2O_5$ (x = 0.03, 0.05) and (1-x)$KNbO_3$.$SiO_2$ (x = 0.1) as a function of frequency in the range from 1 MHz to 1 GHz.

compositions under investigation have large values of dielectric constants. The previously reported poor values have been attributed to the presence of potassium and oxygen vacancies as well as the strains produced in these crystals due to evaporation of potassium and oxygen[13]. Hsiang et *al.* [37] reported that electrical properties are density dependent. In the present work, the $SiO_2$-added composition was poorly densified and hence deteriorated the electrical properties.

## 4 Conclusions

In conclusion, we investigated the effects of 0.03 and 0.05 wt% $P_2O_5$ and 0.1 wt% $SiO_2$ addition on the phase, microstructure and electrical properties of $KNbO_3$. The values of 81.95 and 84.3% of theoretical density were obtained for stoichiometric KN after sintering at 1000 and 1020 °C for 2 h. The TG/DTA analysis of as-mixed-milled sample showed an overall weight loss of 33.4 wt% in the temperature range of 30-1200 °C. The exothermic behavior, observed at about 795 °C, revealed the beginning of crystallization. The addition of low content $P_2O_5$ as a dopant resulted in the formation of an unknown secondary phase and promoted the densification of the material. The microstructure of the samples comprised cubic and cuboidal grains with an average grain size of 3.76 $\mu$m and 3.86 $\mu$m along with agglomerates of fine grains. The grain size increased slightly with an increase in sintering temperature. Dielectric constant and tan$\delta$ were observed to decrease with an increase in x value as well as frequency.

## Acknowledgements

Authors greatly acknowledge the financial support from the Higher Education Commission (HEC) of Pakistan and laboratory support extended by MRL, Department of Physics, University of Peshawar, Pakistan.